# A Be-shell star rotating at the critical limit and a partially stripped companion in a post-mass-transfer solution for the eclipsing binary V505 Mon

Norbert Hauck

Bundesdeutsche Arbeitsgemeinschaft für Veränderliche Sterne e.V. (BAV), Munsterdamm 90, 12169 Berlin, Germany; hnhauck@yahoo.com

**Abstract:** *In this study of the eclipsing shell star binary HD 48914 (V505 Mon) existing photometric and radial velocity data have been completed by an unbiased determination of the companion star's surface temperature and new data from the TESS mission. For the mean effective temperatures in our equatorial view, polar/equatorial radii and masses we get about 16000 K (dimmed), 4.27 / 6.37 Rsun and 7.40 Msun for the Be star, and about 14400 K, 6.19 / 7.17 Rsun and 2.03 Msun for the companion star (a contracting hot subdwarf precursor). Their calculated rotational equatorial velocities both are surprisingly high: 467 km/s for the Be star rotating at its centrifugal limit as well as 131 km/s for its former donor star companion. The Be star is surrounded by a large equatorial decretion disk having a radius of ~65 Rsun and fits into the main sequence of stellar models at a sub-solar metallicity of Z ≈ 0.006.*

V505 Mon (HD 48914) is an eclipsing binary system in a 54 days circular orbit and showing an apparent brightness of about 7.22 *V*mag. It has already been studied by Mayer et al. [1] and Chochol & Mayer [2] at the beginning of this century. However, the evolutionary state as well as the parameter values of its stellar components still remain uncertain, see Rivinius et al. [3]. Therefore, we will try to clarify the situation in our present study.

Mayer et al. [1] and Chochol & Mayer [2] have reported a stationary Hα emission with a line width of about 400 km $s^{-1}$, spectral types from B2 to B8 and luminosity classes from Ib to III. Instead of trying to interprete or disentangle the variable composite spectra of this shell star binary, we prefer to determine first of all the effective temperature $T_{eff}$ of the companion star by *UBV* photometry. The shape of the light curve (see [1] and [2]) is similar to the one encountered in our recent study of the post-mass-transfer eclipsing shell-star binary V658 Car, see Fig. 1 in Hauck [4]. However, the loss of brightness around the secondary minimum is lacking in V505 Mon, and the purely rotational oszillations in the light curve of V658 Car are replaced here by assumed rotationally modulated pulsations of the Be star visible in our Fig. 1 of the TESS light curve and in Fig. 2 of our best-fit residuals for this secondary minimum. The latter has been selected from the TESS data, since showing here a well balanced and lower level of brightness variations than the usual ± 0.06 mag and, therefore, is considered as being suited for the modeling process with the *Binary Maker 3* (BM3) software.

Like in V658 Car, the presence of a large equatorial **decretion disk** of the Be star is visible in the light curve as a funnel-shaped drop of brightness of 0.15 mag in passbands *V* and *B*, and 0.22 mag in *U*, from phase 0.906 to 0.094 (see Table 1).

Moreover, both partial stellar eclipses are clearly visible at phase 0.0 resp. 0.5 and have a duration of 0.029 phase units each. The primary minimum at phase 0 is *exclusively* attributed to the loss of light of the **companion star** being dimmed here by the decretion disk and partially eclipsed by the Be star. The best-fit model in the BM3 delivers also the total light fraction of the companion star in *UBV* (see Table 2). From Table 3 of Wang & Chen [5] we get the interstellar extinction coefficients $A_U$ = 1.584 $A_V$ and $A_B$ = 1.317 $A_V$. Combining the obtained color indexes $(U–B)_0$ and $(B–V)_0$ with Table 5 of Pecaut & Mamjek [6] finally delivered a $T_{eff}$ of 14380 ± 600 K for the companion star in our equatorial view as well as an interstellar extinction $A_V$ of 0.43 mag.

The mean $T_{eff}$ of ≈ 16015 K for the dimmed **Be-shell star** in our equatorial view is obtainable by comparison with its companion in the BM3-Fit of the secondary minimum. The loss of light at primary minimum in passband *V* observed at the eclipse of the companion star by the entire disk allows for the calculation of the dimming effect of half the disk, which is 0.131 *V*mag and equivalent to an increased mean $T_{eff}$ ≈ 17015 K for the undimmed Be star in our equatorial view. Comparison in the BM3 with a sphere of the same surface area and luminosity in *V* gives a mean $T_{eff}$ ≈ 18965 K for the total surface area of this extreme rotator. Finally, with help of Fig. 2, 4 and 5 of Ekström et al. [7] we get a $T_{eff}$ of ≈ 21490 K and a bolometric luminosity of ≈ 3600 $L_\odot$ for the theoretical non-rotating spherical star, which at a mass of 7.40 $M_\odot$ (see here below) fits into the middle of the main sequence of the stellar models of Georgy et al. [8] at evolutionary step 52 and a sub-solar metallicity Z of 0.006.

In our best fit model to the TESS data of the selected secondary minimum the Be-shell star has *extreme* rotational properties (see Table 2), i.e. in the Wilson-Devinney notation a $R_{point}$ /$R_{pole}$ ratio of 1.50, which is the critical resp. centrifugal limit in the Roche model. Therefore, its decretion disk is expected to extend to the Roche lobe limit. In our model, a mass ratio of $M_1/M_2$ = 3.645(5) is then required for reproducing the observed duration of the disk eclipse. Combining this mass ratio with the companion star's radial velocity semiamplitude $K$ = 93.4 km $s^{-1}$ from Mayer et al. [1], the orbital period and inclination $i$ in the mass function delivers the masses $M_1$ = 7.40 $M_\odot$, $M_2$ = 2.03 $M_\odot$ and (with Kepler III) the orbital radius $a$ = 126.57 $R_\odot$. Their error margins have been calculated for an estimated error in $K$ of ± 0.3 km $s^{-1}$.

In analogy to the Be star (see here above) a mean $T_{eff}$ of 14965 K for the total surface area of the significantly slower rotating **companion star** has been derived in the BM3, as well as a $T_{eff}$ of 15770 K and a bolometric luminosity of ≈ 2190 $L_\odot$ for the theoretical non-rotating spherical star with help of Ekström et al. [7]. At a mass of 2.03 $M_\odot$ this companion star does not fit into any known stellar model for post-mass-transfer remnants from 0.765 to 1.95 $M_\odot$ of possible initial donor star masses from 5 to 9.85 $M_\odot$ (see Fig. 1 to 3 of Iben & Tutukov [9]). Therefore, we can assume that the mass transfer has been stopped by the contraction following helium ignition in the donor star, which can happen, if the mass transfer has started close to the tip of the first giant branch. Such an only *partially* stripped remnant then keeps an increased hydrogen mass fraction in its shell. The contraction of a more massive and larger shell towards the hot subdwarf stage should lead to an elevated rotation rate of this remnant, because of the conservation of its angular momentum (also known as the 'pirouette effect'). And indeed, our companion star's rotation rate of ~130 km $s^{-1}$ is clearly higher than the values for normal donor star remnants of ≤ 60 km $s^{-1}$, which have been reported in Table 4 of Rivinius et al. [3].

The **distance** modulus calculation for the companion star having a mean radius of 6.665 $R_{\odot}$ in our equatorial view gives a distance of ~ 819 pc fitting well to the photogeometric distance of 829 (790 – 879) pc from Bailer-Jones et al. [10] for GAIA DR3 data. Moreover, a similar distance modulus calculation for our Be-shell star having a mean radius of 4.94 $R_{\odot}$ in our equatorial view gives ~ 807 pc, which is in line with these results and hence does not indicate any significant amount of disk light in passband *V*.

The very fast rotating Be star implies a spin-up by a recent mass transfer, and is obviously the former accretor in a very young and clearly detached post-mass-transfer binary (see Fig. 3, 4). The fastest currently known rotator among non-degenerate stars appears to be VFTS 102, a massive O-type star formed by a supposed close binary merger in the Large Magellanic Cloud (LMC), with an equatorial rotational velocity of 649 ± 52 km $s^{-1}$ and a critical rotational velocity of 652 km $s^{-1}$ (see Shepard et al. [11]). They derived an equatorial radius $R_e$ of 8.08 $R_{\odot}$ at a polar radius $R_p$ of 5.41 ± 1.55 $R_{\odot}$. Hence the *mean* $R_e$ / $R_p$ ratio thereof is, however only within their larger error margins, more or less comparable with our extreme ratio of 1.50 for V505 Mon having a similar sub-solar LMC-metallicity level of Z ≈ 0.006.

**Table 1: Parameters of the binary system V505 Mon**

| | | |
|---|---|---|
| Epoch [HJD] of phase 0 | 2445253.71(6) | from Vogt & Sterken [12] |
| Orbital period [days] | 53.7745(17) | from Vogt & Sterken [12] |
| Maximum light in *V/B/U* [mag] | 7.22/7.22/6.78 | from Table 1 of Mayer et al. [1] |
| Depth disk eclipse *V/B/U* [mag] | 0.15/0.15/0.22 | from Fig. 2 of Mayer et al. [1] |
| 1. Contact prim. min. *V/B/U* [mag] | 7.37/7.37/7.00 | |
| Mid primary min. *V/B/U* [mag] | 7.58/7.58/7.18 | |
| Depth second. min. *V/B/U* [mag] | 0.32/0.32/0.24 | from Table 1 of Mayer et al. [1] |
| Eclipse duration [days] | 10.11 | disk eclipse |
| Eclipse duration [days] | 1.56 | stellar eclipses |
| Orbital inclination $i$ [deg] | 87.89 ± 0.03 | and circular orbit adopted |
| Orbital radius $a$ [$R_\odot$] | 126.57 ± 0.60 | for $R_\odot$ = 696342 km |
| Distance [pc] | 819 ± 31 | photometric |
| Extinction $A_V$ [mag] | 0.433 | interstellar |

**Table 2: Parameters of the components of V505 Mon**

| Parameter | Be star | Companion star | Disk |
|---|---|---|---|
| Radius R (mean) [$R_\odot$] | 5.67 ± 0.10 | 6.85 ± 0.14 | 64.7 ± 1.6 |
| Radius R (pole/equator/point) [$R_\odot$] | 4.27/6.37/6.39 | 6.19/7.17/7.18 | |
| $R_{point}/R_{pole}$ ratio | 1.50 (+0/–0.02) | 1.16 ± 0.03 | |
| Rotational velocity [km $s^{-1}$] | 467 (+1/–8) | 131 ± 13 | |
| Rotational period [days] | 0.691 | 2.779 | |
| Mean $T_{eff}$ [K] (incl. disk in our view) | 16015 ± 670 | 14385 ± 600 | |
| *V*-flux fraction at max. light | 0.398 | 0.602 | 0 adopted |
| *B*-flux fraction at max. light | 0.398 | 0.602 | |
| *U*-flux fraction at max. light | 0.431 | 0.569 | |
| Apparent *V* magnitude | 8.220 | 7.771 | |
| Apparent *B* magnitude | 8.220 | 7.771 | |
| Apparent *U* magnitude | 7.693 | 7.393 | |
| Luminosity (bolometric; no disk, in our view) [$L_\odot$] | ~1836 | ~1706 | |
| Mass [$M_\odot$] | 7.40 ± 0.07 | 2.03 ± 0.02 | |

**Acknowledgements :**
This research includes data collected with the TESS mission, obtained from the MAST data archive at the Space Telescope Science Institute (STScI). Moreover, it has used the Simbad and VizieR databases operated at the *C*entre de *D*onnées astronomique, *S*trasbourg, France.

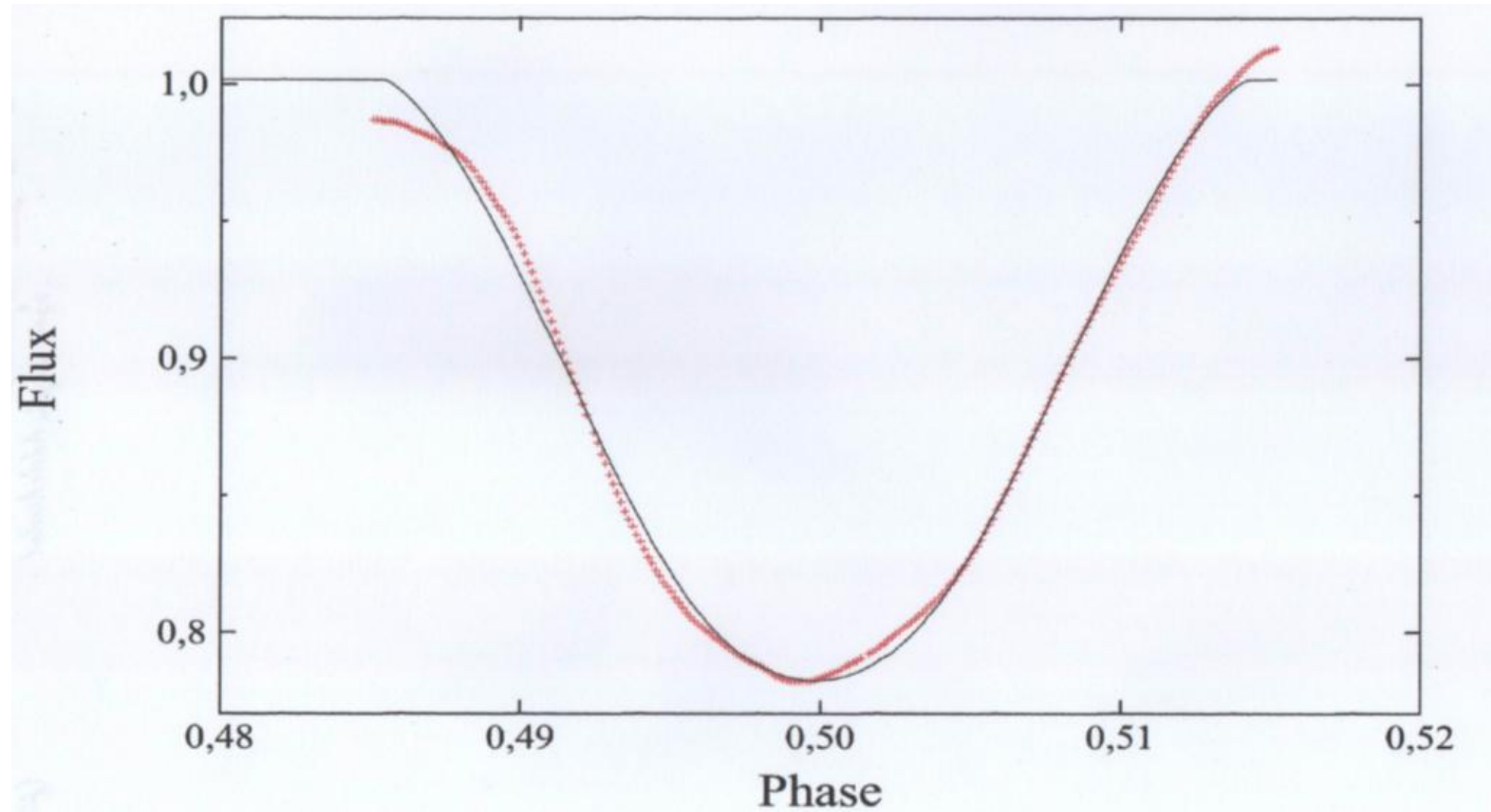

**Fig. 1:** Light curve fit ($\sigma_{FIT}$ = 5.9 mmag) to 232 TESS data points (red crosses) of the secondary minimum's partial eclipse in December 2020.

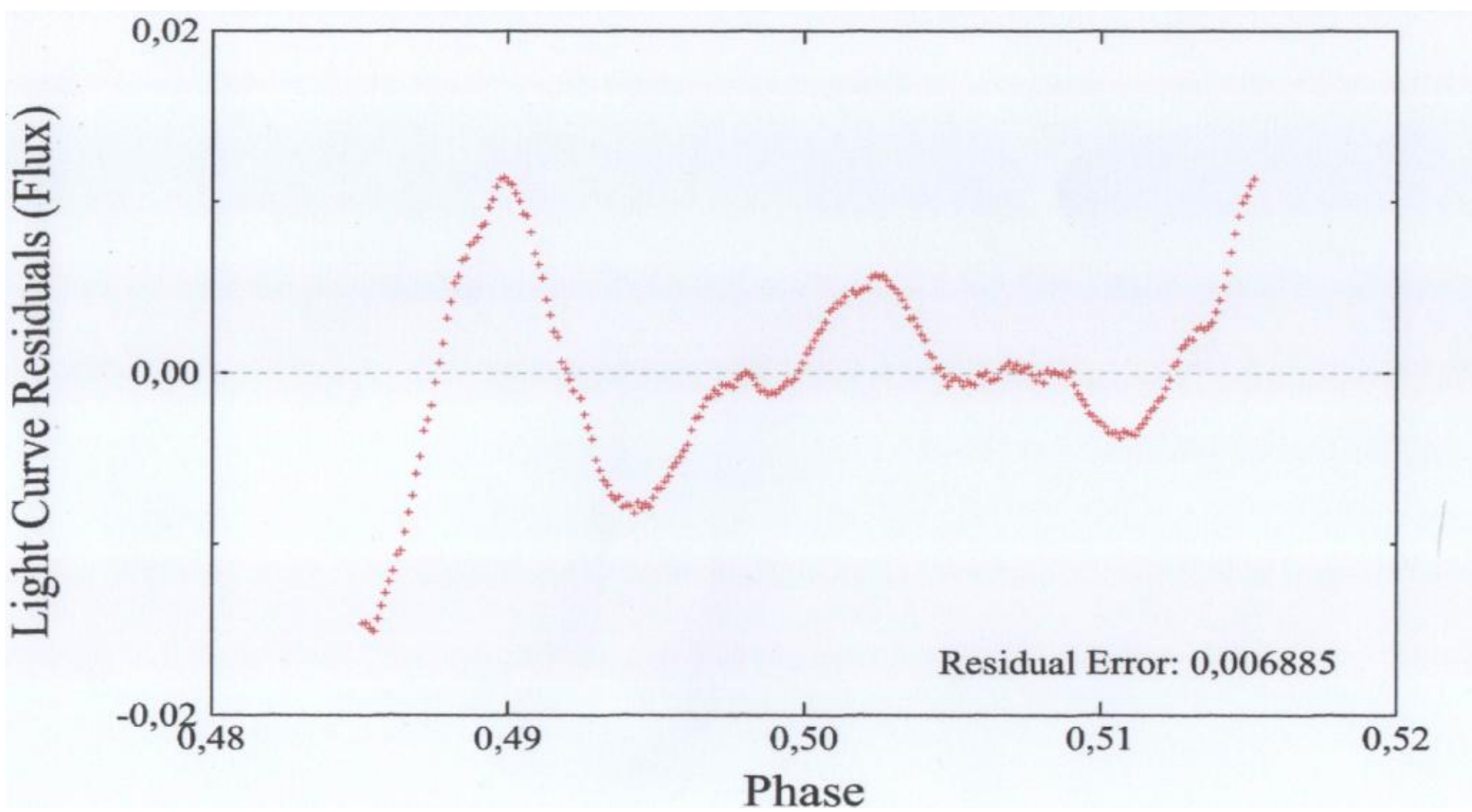

**Fig. 2** showing *simultaneously* the minimized light curve residuals error of Fig. 1 and a nearly perfect fit for the end of the ingress and the middle of the egress, when the shape of the Be star is obviously not being deformed by pulsations.

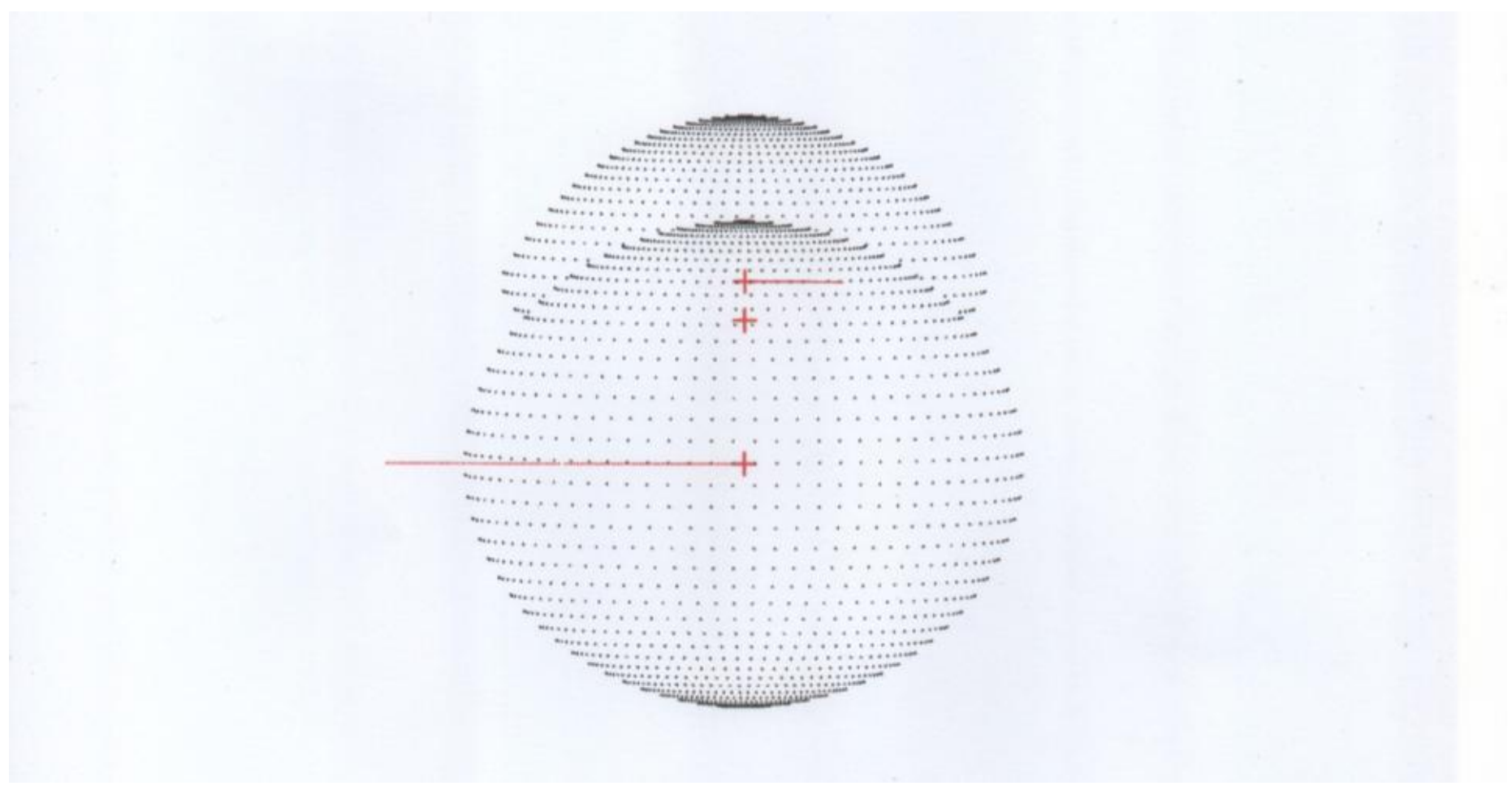

**Fig. 3:** Partial eclipse of the Be star by its slightly larger companion star in the middle of the secondary minimum (phase 0.5).

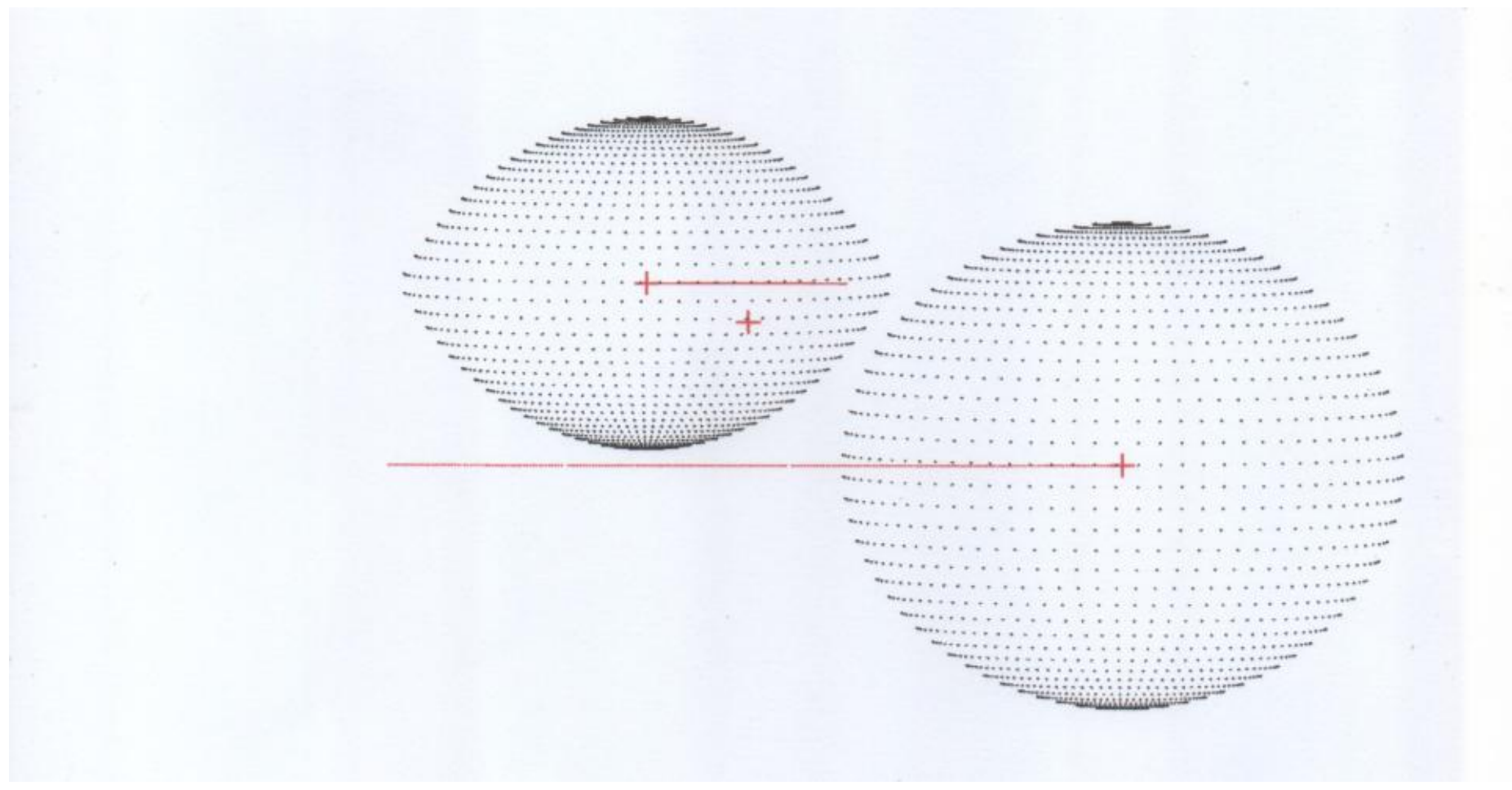

**Fig. 4:** Shortly after the secondary minimum: Note the oblate spheroidal shape of the slower rotating companion star and the biconvex-lens shape of the Be star rotating at its centrifugal resp. critical limit.